\documentclass[prl,aps,twocolumn,
pdflatex,
superscriptaddress,floats,showpacs,10pt]{revtex4-1}
\usepackage{graphicx}
\usepackage{dcolumn}
\usepackage{bm}
\usepackage{amssymb}
\usepackage{amsmath}
\usepackage{textcomp}
\usepackage{color}

\begin{document}

\title{Effect of Electromagnetic Pulse Transverse Inhomogeneity\\
on the Ion Acceleration by Radiation Pressure}

\author{K. V. Lezhnin}
\affiliation{Moscow Institute of Physics and Technology, 
Institutskiy per. 9, Dolgoprudny, Moscow Region, 141700,
Russia}

\author{F. F. Kamenets}
\affiliation{Moscow Institute of Physics and Technology, 
Institutskiy per. 9, Dolgoprudny, Moscow Region, 141700,
Russia}

\author{V. S. Beskin}
\affiliation{Moscow Institute of Physics and Technology, 
Institutskiy per. 9, Dolgoprudny, Moscow Region, 141700,
Russia}
\affiliation{Russian Acad. Sci., P. N. Lebedev Phys. Inst., Leninskii Prosp 53, Moscow 119991, Russia}

\author{M. Kando}
\affiliation{Japan Atomic Energy Agency, Kansai Photon Science Institute, 
8-1-7 Umemidai, Kizugawa-shi, Kyoto, 619-0215 Japan}

\author{T. Zh. Esirkepov}
\affiliation{Japan Atomic Energy Agency, Kansai Photon Science Institute, 
8-1-7 Umemidai, Kizugawa-shi, Kyoto, 619-0215 Japan}

\author{S. V. Bulanov}
\altaffiliation{Also at the ITMO University, Saint-Petersburg 197101, Russia; Russian Acad. Sci., A. M. Prokhorov General Phys. Inst., 
Vavilov Str. 38, Moscow, 119991, Russia}
\affiliation{Moscow Institute of Physics and Technology, 
Institutskiy per. 9, Dolgoprudny, Moscow Region, 141700,
Russia}
\affiliation{Japan Atomic Energy Agency, Kansai Photon Science Institute, 
8-1-7 Umemidai, Kizugawa-shi, Kyoto, 619-0215 Japan}

\date{\today}

\begin{abstract}
In the ion acceleration by radiation pressure a transverse inhomogeneity of the electromagnetic pulse 
results in the displacement of the irradiated target in the off-axis direction limiting achievable ion energy. 
This effect is described analytically within the framework of the thin foil target model and with 
the particle-in-cell simulations showing that the maximum energy of accelerated ions decreases 
while the displacement from the axis of the target initial position increases. 
The results obtained can be applied for optimization of the ion acceleration 
by the laser radiation pressure with the mass limited targets.

\bigskip


\end{abstract}

\maketitle

\section{Introduction}

Studies of the high energy ion generation in the interaction between an ultraintense laser
pulse and a small overdense targets, are of
fundamental importance for various research fields ranging
from the developing the ion sources for
thermonuclear fusion and medical applications to the investigation of high energy density
phenomena in relativistic astrophysics (see review articles \cite{IonsB, IonsAVK, IonsD, IonsM, IonsGUS, HTUFN, RelAstro} and 
the literature cited therein). 

Theory and experiments on laser acceleration can clarify the basic features 
of particle acceleration in astrophysical objects. Indeed, according to common point of view, 
activity of radio pulsars, active galactic nuclei, and even gamma-bursters connects 
with the highly magnetized wind in which electric field is approximately equal 
to magnetic one \cite{BesBook}. Charged particles produced in such a wind can get 
the energy of the order of $m_{\alpha} c^2 \gamma_w^2$ that is much higher than the energy of 
the wind ($\approx m c^2 \gamma_w$) \cite{Load1, Load2}. Here $\gamma_w$ is the Lorentz-factor 
associated with the wind velocity. Moreover, interacting with the external 
environment (the companion star in a close binary system, the current sheets in the pulsar wind),  
a region where the electric field is greater than the magnetic can form where therefore 
the acceleration of particles can be even more effective \cite{Prince}.

\begin{figure}
\centering
\includegraphics[width=6 cm]{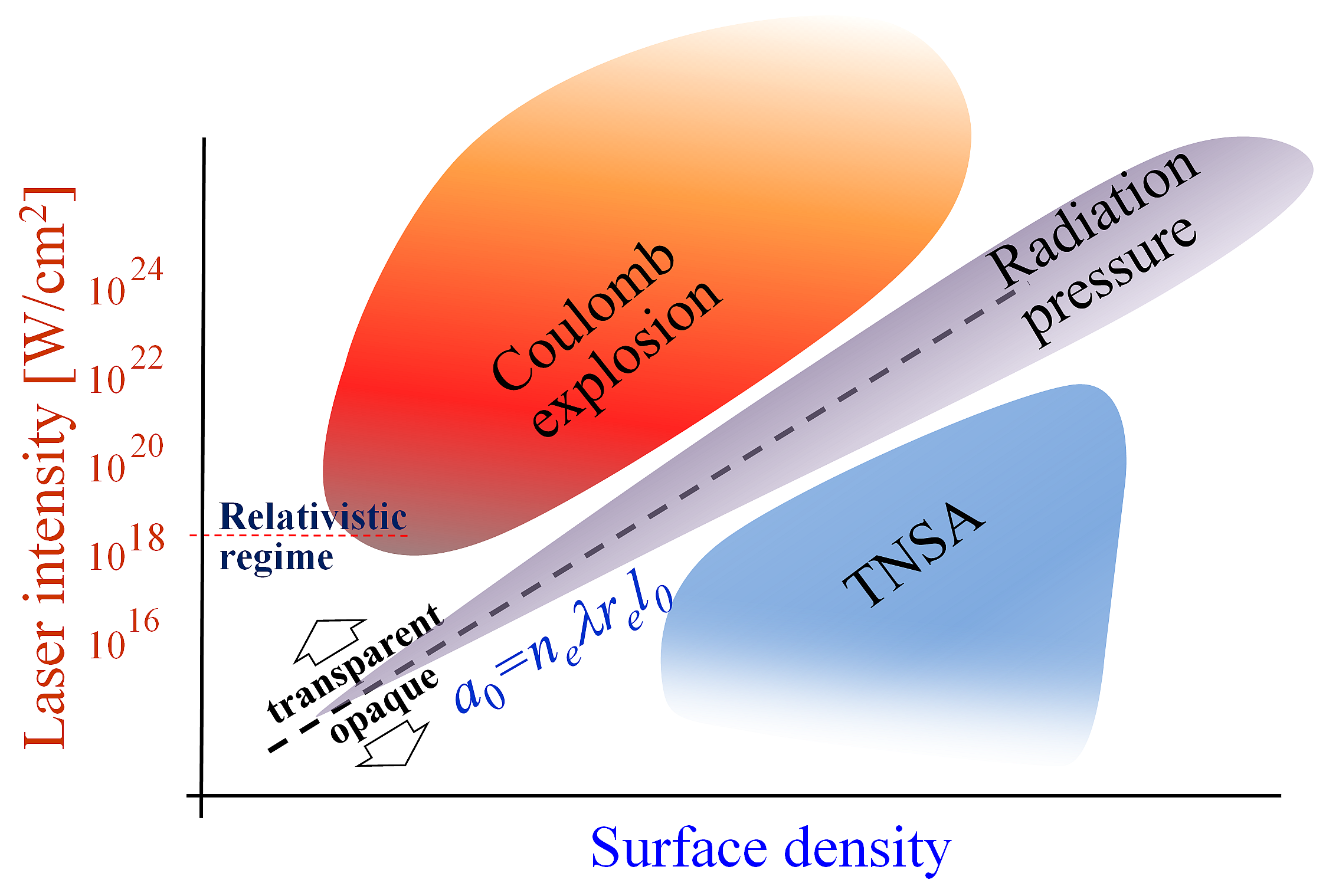}
\caption{The ion acceleration 
regimes in the plane of the laser intensity [W/cm$^2$]--the surface density $n_e l_0$ of the target.}
\label{fig:IonAccRegimes}
\end{figure}

Depending on the laser and target parameters different regimes of acceleration appear -- 
from acceleration at the target surface called the Target Normal Sheath Acceleration (TNSA) \cite{TNSAG, TNSAW, TNSAH} 
through the Coulomb explosion \cite{CEL, CEN, CEB, MM} 
to radiation pressure dominance acceleration (RPDA) regime \cite{RPDAV, RPDAE, RPDAK, RPDAR}. The ion acceleration 
regimes are shown in the plane of the laser intensity -- the surface density $n_e l_0$ of the target in 
Fig. \ref{fig:IonAccRegimes} (see also \cite{HTUFN}). Here $n_e$ is the electron density in the target 
and $l_0$ is its thickness. 
At the intensity above $10^{18}{\rm W/cm^2}$ the plasma electron energy becomes relativistic.  The dashed line,
is given by the formula 
\begin{equation}  
\label{eq:tr-op}
a_0=n_e r_e \lambda l_0,
\end{equation}
where $a_0=eE_0/m_e \omega c$ 
is the normalized laser pulse amplitude,
$\omega$ and $\lambda=2\pi c/\omega$ are the laser frequency and wavelength, respectively, 
and $r_e=e^2/m_ec^2=2.8\times 10^{-13}$cm is the classical electron radius; $m_e$ and $e$ 
are the electron mass and charge, and 
$c$ is the speed of light in vacuum. This line separates the intensity -- surface density plane into two domains. 
In the domain below the line the plasma is opaque and above it is transparent for the laser radiation \cite{FOILV}. 
When the laser radiation interacts with the opaque target a relatively small portion of hot electrons 
can escape forming 
a sheath with strong electric charge separation electric field where the acceleration occurs in the TNSA regime. 
Above the dashed line the laser radiation is so intense that it blows out almost all electrons from 
the target irradiated region. The remaining ions undergo fast expansion, the Coulomb explosion, due repelling 
of noncompensated electric charges. At the opaqueness-transparency threshold, in the vicinity of the dashed line in 
Fig. \ref{fig:IonAccRegimes} the optimal conditions for the ion acceleration in the RPDA regime 
are realized \cite{RPDAE, RPDASS}. A fundamental feature of
the RPDA acceleration process, proposed by Veksler \cite{RPDAV}, is its high efficiency, as the ion
energy per nucleon turns out to be proportional in the ultrarelativistic
limit to the electromagnetic pulse energy. As far it concerns the experimental 
evidence of the RPDA mechanism there are indications on its realization in the laser thin foil interaction 
reported in Refs. \cite{SKar1, Henig, SKar2}.

The usage of a finite transverse size target, it is called the Mass Limited Target (MLT) or the Reduced Mass Target 
\cite{JLIM, AAA, Kluge, Zeil, Yu, AZ, Wang}, including the cluster targets 
\cite{CEN, Fukuda2009}, provide a way for enhancement of the ion energy and acceleration efficiency and a way for 
high brightness X-ray generation \cite{YuPRL}. 
The irradiation of MLT by enough high intensity lasers 
is one of the most perspective approaches to develop compact ion accelerators \cite{UNLIM, UFNMIRR}. 

In the present paper, we discuss the RPDA regime under the conditions when a transverse inhomogeneous 
laser pulse irradiates a MLT positioned not precisely at the laser pulse axis. This situation natually occurs 
due to a finite pointing stability of the laser systems.
As a result the transverse component of the radiation pressure leads to the displacement of the irradiated target 
in the off-axis direction. Apparently, after a finite interval of time the target leaves the laser pulse 
preventing from the further ion acceleration. Below on the ground of a theoretical model 
of relativistic mirror \cite{RPDAE, UNLIM, UFNMIRR} we calculate the acceleration time and hence the achieved 
ion energy dependence on the laser pulse amplitude and transverse size and on the initial displacement of the target 
from the laser axis. According to recently published papers, various instabilities of the target plasma 
appear in the RPDA regime, for instance, the Rayleigh-Taylor-like instability \cite{PhotonBubbles} 
leads to the target modulation forming the low density bubbles and high density clumps resulting 
in the broadening of the accelerated ion energy spectrum. In order to elucidate the kinetic, nonlinear and 
instability effects we carry out the PIC simulations of the finite waist laser pulse interaction 
with the MLT by using the REMP code \cite{REMP}. 

\section{Dynamics of the Mass Limited Target Positioned Slightly Off-Axis}

\subsection{The Equations of Motion}
We describe the nonlinear dynamics of a laser accelerated target
within the framework of the thin shell approximation
formulated by Ott \cite{Ott} and further generalized on the 3D geometry 
in Refs. \cite{gen1, gen2} and 
extended to the relativistic case in Refs. \cite{UNLIM, PhotonBubbles}.

In a way of Refs. \cite{UNLIM, UNLIMPoP, PhotonBubbles} here we derive 
of the motion equations required for further consideration of the MLT dynamics.
The equations of motion of the surface element of a thin foil target 
in the laboratory frame of reference can be
written in the form 
\begin{equation}  
\label{eq1}
\frac{d \mathbf{p}}{dt} = \frac{\mathcal{P} \mathbf{\nu} }{\sigma },
\end{equation}
\noindent where $\mathbf{p}$, $\mathcal{P}$, $\mathbf{\nu}$, and $\sigma $
are the momentum, light pressure, unit vector normal to the shell surface
element, and surface density, $\sigma = nl_0$, respectively. Here $n$ and $l$
are the plasma ion density and shell thickness.
We determine the surface element $\Delta s$ as carrying 
$\Delta N = \sigma \Delta s$ particles, with $\Delta N$ constant in time. We take
the shell initially to be at rest, at $t = 0$, in the plane $x = 0$. In
order to describe how its shape and position change with time it is
convenient to introduce the Lagrange coordinates $\eta $ and $\zeta $
playing the role of the markers of the shell surface element. The shell
shape and position are given by the equation 
\begin{equation}  
\label{eq2}
\mathbf{M} = \mathbf{M}(\eta ,\zeta ,t) \equiv \{x(\eta ,\zeta ,t),
y(\eta,\zeta ,t),z(\eta ,\zeta ,t)\}.
\end{equation}
At a regular point, the surface area of a shell element and the unit vector
normal to the shell are equal to
\begin{equation}  
\label{eq3}
\mathbf{\nu} \Delta s = \partial _\eta \mathbf{M}\times \partial _\zeta 
\mathbf{M}\,\mbox{d}\eta \mbox{d}\zeta
\end{equation}
\noindent and
\begin{equation}  
\label{eq4}
\mathbf{\nu} = \frac{\partial _\eta \mathbf{M}\times \partial _\zeta 
\mathbf{ M}} {\vert \partial _\eta \mathbf{M}\times \partial _\zeta \mathbf{M}\vert },
\end{equation}
respectively (see e.g., \cite{KndK}). The particle number
conservation implies $\sigma \Delta s = \sigma _0 \Delta s_0 $, where 
$\sigma _0 = n_0 l_0 $. This yields
\begin{equation}  
\label{eq5}
\sigma = \frac{\sigma _0 }{\vert \partial _\eta \mathbf{M}\times \partial
_\zeta \mathbf{M}\vert }.
\end{equation}
Using these relationships and representing the coordinates $x_i$ as 
\begin{gather}
x=\xi_x(\eta,\zeta,t),\\
y=\eta+\xi_y(\eta,\zeta,t),\\ 
z=\zeta+\xi_z(\eta,\zeta,t)
\end{gather}
with initial conditions: $\xi_i(\eta,\zeta,0)=0$ and $\dot\xi_i(\eta,\zeta,0)=v_i(\eta,\zeta,0)$, 
we obtain the equations of motion 
in the form \cite{Ech10}
\begin{gather} 
\label{eq:mot-px}
\sigma _0 \partial _t p_x = 
\mathcal{P} \left(1+ \partial _\eta \xi_y + \partial _\zeta \xi_z +\{\xi_y,\xi_z\} \right),\\
%
\label{eq:mot-py}
\sigma _0 \partial _t p_y = 
\mathcal{P} \left(- \partial _\eta \xi_x +\{\xi_z,\xi_x\} \right),\\
%
\label{eq:mot-pz}
\sigma _0 \partial _t p_z = 
\mathcal{P} \left(- \partial _\zeta \xi_x +\{\xi_x,\xi_y\} \right),\\
\label{eq:mot-xi}
\partial _t \xi_i = c\frac{p_i }{(m_\alpha ^2 c^2 + p_k p_k )^{1 / 2}},
\end{gather}
Here $m_\alpha $ is the ion mass, $i = 1,2,3$, and summation over repeated indices is
assumed, 
\begin{equation}
\{\xi_j,\xi_k\}=\partial _\eta \xi_j\partial _\zeta \xi_k-\partial _\zeta \xi_j\partial _\eta \xi_k
\end{equation}
are Poisson's brackets. This form of the equations is particularly 
convenient for analysing small but finite displacement of the target elements from the axis.

The radiation pressure on the shell exerted by a circularly polarized
electromagnetic wave propagating along the $x$-axis with amplitude 
$E=E(t-x/c)$ is 
\begin{equation}
\mathcal{P}=K\frac{E^{2}}{4\pi }\left( {\frac{1-\beta_x }{1+\beta_x }}\right),
\label{eq8}
\end{equation}
where $\beta_x =p_{x}(m_{\alpha }^{2}c^{2}+p_{x}^{2})^{-1/2}$ is the shell
normalized velocity in the $x$-direction. The coefficient $K$ equal to 
\begin{equation}
K=2|\rho|^2+|\alpha|^2
\end{equation}
depends on $|\rho|^2$ and $|\alpha|^2$ which are the light reflection and absorption coefficients,
respectively (see also Ref. \cite{Slow}). Effects of the reflection coefficient dependence  
on the ion energy due to the relativistic transparency has been discussed in Refs. \cite{RPDASS, OPTAM}.
Below we shall not consider the relativistic transparency effects assuming ideally 
reflecting light target with $K=2$.

We note here that in Eqs. (\ref{eq:mot-px}--\ref{eq:mot-pz}) there is no a force acting between 
the target surface elements, i. e. we can consider a finite transverse 
size MLT for which the Lagrange coordinates $\eta$ and $\zeta$ belong to a finite domain: 
$\eta \in [\eta_{1},\eta_{2}]$ and $\zeta \in [\zeta_{1},\zeta_{2}]$.

For homogeneous laser pulse, $E=$constant, the flat MLT is accelerated along the 
$x$-axis with $p_y=0$, $p_z=0$, $\xi_y=0$, and $\xi_z=0$.
The ion momentum and displacement in the $x$-direction are given by dependences on time \cite{RPDAE}:
 \begin{equation}  
\label{eq:px(0)}
p^{(0)}_x(t) = m_{\alpha}c\left(\frac{t}{t_{1/3}}\right)^{1/3},
\end{equation}
 \begin{equation}  
\label{eq:x(0)}
\xi^{(0)}_x(t) = c t -3 c t_{1/3}^{2/3} t^{1/3},
\end{equation}
where the characteristic time is 
 \begin{equation}  
\label{eq:t{1/3}}
t_{1/3}=\frac{8\pi \sigma_0 m_{\alpha} c}{3 E^2}.
\end{equation}
Here we have assumed that the target energy is ultrarelativistic, $p^{(0)}_x/m_{\alpha}c \gg 1$.

\begin{figure}
\centering
\includegraphics[width=7.5 cm]{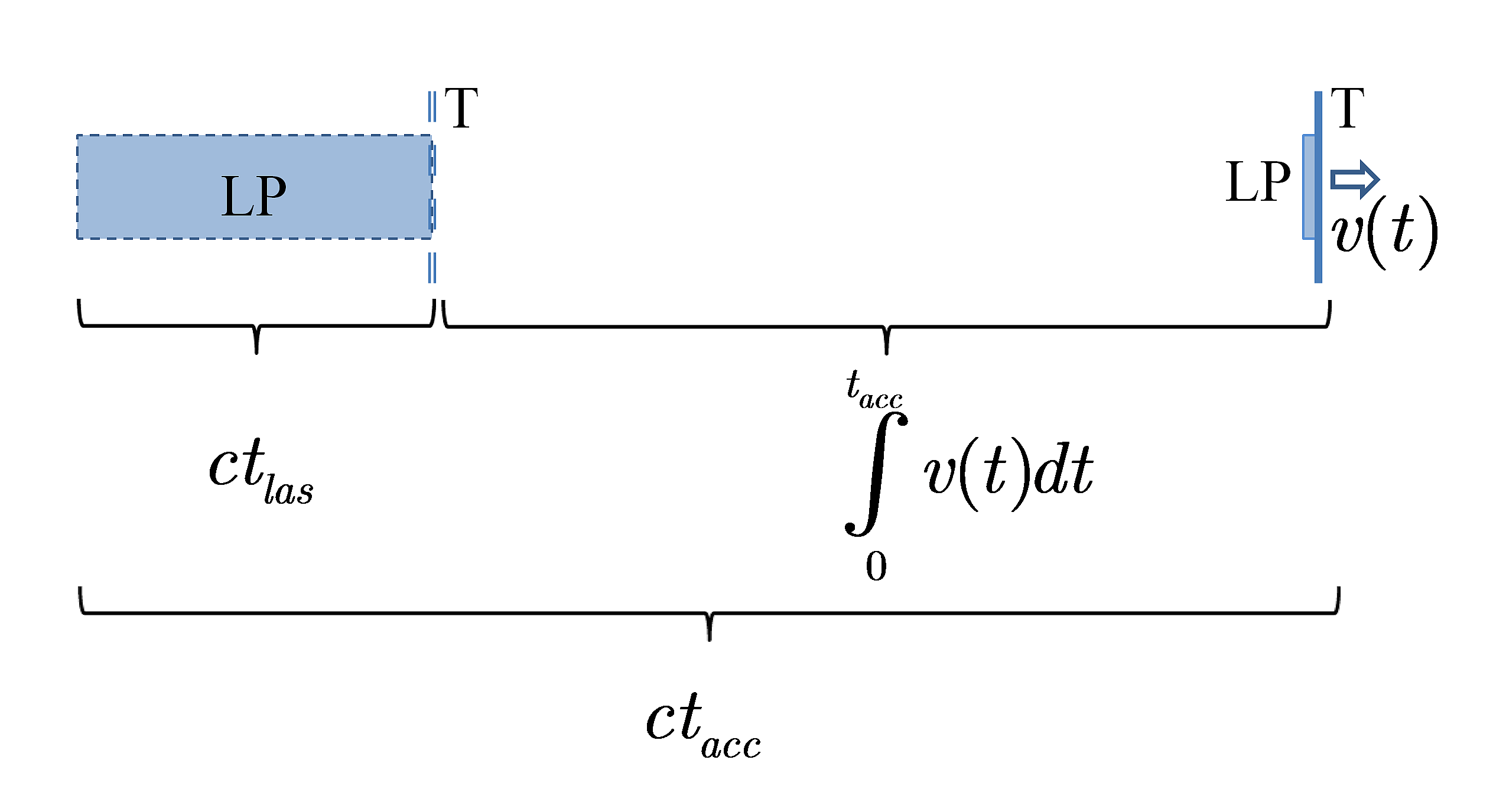}
\caption{Laser pulse $LP$ and the target $T$ at $t=0$ and $t=t_{acc}$}
\label{Figtacc}
\end{figure}

Using relationships (\ref{eq:px(0)}) and (\ref{eq:x(0)}) we can easily find that the finite duration, $t_{las}$,
laser pulse accelerates the ions up to the energy ${\cal E}=m_{\alpha} c^2 \gamma_{max}$ with the gamma-factor 
given by 
 \begin{equation}  
\label{eq:gamma(0)}
\gamma_{max} = \frac{E^2 t_{las}}{4\pi \sigma_0 m_{\alpha} c}.
\end{equation}
According to Eq. (\ref{eq:px(0)}) the acceleration time, $t_{acc}$, can be defined 
via $\gamma_{max}= (t_{acc}/t_{1/3})^{1/3}$. We find it taking into account that the 
laser pulse rear reaches the target at $t=t_{acc}$, as it is illustrated in Fig. \ref{Figtacc}.
The acceleration time is determined by equation
 \begin{equation}  
\label{eq:vtacc}
t_{las} =\int^{t_{acc}}_0\left(1-\frac{v(t)}{c} \right)dt 
\approx \frac{1}{2}\int^{t_{acc}}_0\frac{dt}{\gamma(t)^2}dt
\end{equation}
This and Eq. (\ref{eq:gamma(0)}) yield 
 \begin{equation}  
\label{eq:t_acc}
t_{acc} =\frac{2}{3} \gamma_{max}^2t_{las}.
\end{equation}

\subsection{The Mass Limited Target Irradiated by Gaussian Laser Pulse}
In order to analyse the transverse motion of the MLT irradiated by the laser pulse we consider the pulse 
whose envelope has a Gaussian form,
 \begin{equation}  
\label{eq:Gauss}
E(y,z) =E_0 \exp\left(-\frac{y^2}{2 l_y^2}-\frac{z^2}{2 l_z^2} \right)
\end{equation}
with the laser pulse width equal to $l_y$ and $l_z$ in the $y$- and $z$-direction, respectively.

Assuming a smallness of the transverse displacement, $\xi_y \ll \eta$, $\xi_z \ll \zeta$, and considering 
the near-axis region, $\eta \ll l_y$, $\zeta \ll l_z$, we obtain from Eqs. (\ref{eq:mot-px}--\ref{eq:mot-pz}) 
the linearized system of equations,
\begin{gather} 
\partial _t \left((\gamma^{(0)} (t))^3 \partial _t \xi_x^{(1)}\right) = \nonumber \\
\frac{c}{(\gamma^{(0)} (t))^2 t_{1/3}^{(0)}}
 \left(\partial _\eta \xi_y^{(1)} + \partial _\zeta \xi_z^{(1)} -
 \frac{\eta^2}{l_y^2} -\frac{\zeta^2}{l_z^2} \right),\label{eq:mot-x1} \\
\label{eq:mot-y1}
\partial _{t} \left(\gamma^{(0)} (t) \partial _t \xi_y^{(1)}\right) = -\frac{c}{(\gamma^{(0)} (t))^2 t_{1/3}^{(0)}}
 \partial _\eta \xi_x^{(1)},\\
\label{eq:mot-z1}
\partial _{t} \left(\gamma^{(0)} (t) \partial _t \xi_z^{(1)}\right) = -\frac{c}{(\gamma^{(0)} (t))^2 t_{1/3}^{(0)}}
 \partial _\zeta \xi_x^{(1)}
\end{gather}
 with given dependence on time of the ion gamma-factor
  \begin{equation}  
\label{eq:gamma0}
\gamma^{(0)}(t)=\left(\frac{t}{t_{1/3}^{(0)}}\right)^{1/3},
 \end{equation} 
 which is found within the framework of the 1D model of the RPDA thin foil acceleration \cite{RPDAE}. 
 The approach used corresponds to so-called betatron approximation well known in the theory of standard accelerators 
 of charged particles \cite{Humphries}. 
 In these expressions the characteristic time is $t_{1/3}^{(0)}=8\pi \sigma_0 m_{\alpha} c/3 E_0^2$.
 
 In order to find the solution to the system of equations in partial derivatives 
 (\ref{eq:mot-x1}--\ref{eq:mot-z1}) we use the anzatz
 \begin{gather}
 \xi_x^{(1)}(\eta,\zeta,t)=\Xi_x(t)+\Xi_{x\eta\eta}(t)\eta^2+\Xi_{x\zeta\zeta}(t)\zeta^2,\\
 \xi_y^{(1)}(\eta,\zeta,t)=\Xi_{y\eta}(t)\eta, \label{eq:xi-perp}\\
 \xi_z^{(1)}(\eta,\zeta,t)=\Xi_{z\zeta}(t)\zeta,
 \end{gather}
 which is a self-similar solution reducing Eqs. (\ref{eq:mot-x1}--\ref{eq:mot-z1}) to
 ordinary differential equations for the functions $\Xi_x(t)$, $\Xi_{x\eta\eta}(t)$,
 $\Xi_{x\zeta\zeta}(t)$, $\Xi_{y\eta}(t)$, and $\Xi_{z\zeta}(t)$:
 \begin{gather}
 \label{eq:Xixyy1}
\frac{d}{d\tau} \left(\gamma^{(0)} \frac{d\, \Xi_{x\eta\eta}}{d\tau}\right) = -\frac{1}{l_y^2},\\
 \label{eq:Xixzz1}
\frac{d}{d\tau} \left(\gamma^{(0)} \frac{d\, \Xi_{x\zeta\zeta}}{d\tau}\right) = -\frac{1}{l_z^2},\\
\label{eq:Xiyy1}
\frac{d}{d\tau} \left(\frac{1}{\gamma^{(0)}} \frac{d\, \Xi_{y\eta}}{d\tau}\right) = -2 \Xi_{x\eta\eta},\\
\label{eq:Xizyy1}
\frac{d}{d\tau} \left(\frac{1}{\gamma^{(0)}} \frac{d\, \Xi_{z\zeta}}{d\tau}\right) = -2 \Xi_{x\zeta\zeta},\\
\label{eq:Xix1}
\frac{d}{d\tau} \left(\gamma^{(0)} \frac{d\, \Xi_{x}}{d\tau}\right) = \Xi_{y\zeta}+\Xi_{z\zeta}.
 \end{gather}
 We introduced a new independent variable equal to
 \begin{equation}  
\label{eq:tau}
\tau =\left(\frac{c}{t_{1/3}^{(0)}}\right)^{1/2}\int_0^t\frac{dt}{(\gamma^{(0)} (t))^2}
\approx 3 c^{1/2} (t_{1/3}^{(0)})^{1/6}t^{1/3}.
\end{equation}
For initial conditions $\xi_x^{(1)}(\eta,\zeta,0)=0$ and $\dot\xi_x^{(1)}(\eta,\zeta,0)=0$, 
solution to Eqs. (\ref{eq:Xixyy1}--\ref{eq:Xix1}) reads
 \begin{gather} 
 \Xi_{x\eta\eta}=-\frac{9 c t}{l_y^2}\left(\frac{t_{1/3}}{t}\right)^{2/3}, \,
 \Xi_{x\zeta\zeta}=-\frac{9 c t}{l_z^2}\left(\frac{t_{1/3}}{t}\right)^{2/3},\\
 \Xi_{y\eta}=\frac{81 (c t)^2}{4\,l_y^2}\left(\frac{t_{1/3}}{t}\right)^{2/3}, \,
 \Xi_{z\zeta}=\frac{81 (c t)^2}{4\,l_z^2}\left(\frac{t_{1/3}}{t}\right)^{2/3}, \label{eq:sol}\\
 \Xi_{x}=-\frac{729 (c t)^3}{100}\left(\frac{t_{1/3}}{t}\right)^{4/3}\left(\frac{1}{l_y^2} + \frac{1}{l_z^2}\right).
 \end{gather}

As it is seen from Eqs. (\ref{eq:xi-perp}) and (\ref{eq:sol}) the target element with initial coordinates 
$\eta$ and $\zeta$ moves in the transverse direction with the displacement proportional to $t^{4/3}$. We can 
estimate the time required to leave the region with strong laser field as
 \begin{equation}  
\label{eq:tau-perp}
\delta t_{\perp}=\left(\frac{4 l_{\perp}^3}{81 \delta r_0}\right)^{3/4}\frac{1}{c^{3/2}t_{1/3}^{1/2}}
\end{equation}
with $l_{\perp}=\min\{l_y,l_z\}$ and $\delta r_0=\max\{\eta,\zeta\}$. According to Eqs. (\ref{eq:gamma(0)}) and 
(\ref{eq:tau-perp})  the achieved ion energy is of the order of 
 \begin{equation}  
\label{eq:gamma-perp}
{\cal E}_{\alpha}=m_{\alpha} c^2 \left(\frac{\delta t_{\perp}}{t_{1/3}}\right)^{1/3},
\end{equation}
which implies $\delta t_{\perp} < t_{acc}$. The opposite case realized for small enough initial position of the MLT 
centroid, $\delta r_0$, and/or wide enough laser pulse corresponds to the perfect laser-target alignment.

Using obtained above relationships we can write the characteristic time $t_{1/3}$ as 
 \begin{equation}  
\label{eq:t-1/3}
t_{1/3}=\frac{2 \omega_{pe}^2}{\omega^2} \frac{m_{\alpha}}{m_e}\frac{l_0}{c}\frac{1}{a_0^2},
\end{equation}
which for the solid density target, $\omega_{pe}^2/\omega^2\approx 10^2$, of the thickness $l_0=0.1\mu$m for the 
laser intensity of the order of $10^{23}$W/cm$^2$ corresponding to $a_0=300$, 
$m_{\alpha}=m_p$, yields $t_{1/3}\approx 1.5\,$fs. In the case of perfect laser-target alignment the 
maximal achievable ion (proton) energy $m_{\alpha} c^2(t_{las}/3 t_{1/3})$ for 100 fs laser pulse duration is about 
20 GeV with the acceleration time given by Eq. (\ref{eq:t_acc}) equal to 10 ps. The perfect alignment condition 
implies $t_{\perp}>t_{acc}$.

\subsection{Super Gaussian Laser Pulse Interaction with Mass Limited Target}
Here we analyse the case when the laser pulse when its envelope has Super-Gaussian form,
 \begin{equation}  
\label{eq:Gauss}
E(y,z) =E_0 \exp\left(-\frac{y^4}{2 l_y^4} \right),
\end{equation}
with the index equal to 4. For the sake of brevity we consider two-dimensional geometry.
Generalization to the 3D case is straightforward.

For small transverse displacement, $\xi_y \ll \eta$, in
the near-axis region, $\eta \ll l_y$, within the framework of the betatron approximation 
the target dynamics is described by the linearized system of equations,
\begin{gather} 
\partial _{\tau} \left(\gamma^{(0)} \partial _{\tau} \xi_x^{(1)}\right) = \partial _\eta \xi_y^{(1)}  -
 \frac{\eta^4}{l_y^4},\label{eq:sg-mot-x1} \\
\label{eq:sg-mot-y1}
\partial _{\tau} \left(\frac{1}{\gamma^{(0)}} \partial _{\tau} \xi_y^{(1)}\right) = \partial _\eta \xi_x^{(1)}
\end{gather}
 with the independent variable $\tau$ defined by Eq. (\ref{eq:tau}) and 
 the ion gamma-factor $\gamma^{(0)}$ given by Eq. (\ref{eq:gamma0}).
 
 The self-similar solution to Eqs. (\ref{eq:sg-mot-x1}--\ref{eq:sg-mot-y1}) has a form
 \begin{gather}
 \xi_x^{(1)}(\eta,\tau)=\Xi_x(\tau)+\Xi_{x\eta\eta}(\tau)\eta^2+\Xi_{x\eta\eta\eta\eta}(\tau)\eta^4,\\
 \xi_y^{(1)}(\eta,\tau)=\Xi_{y\eta}(\tau)\eta+\Xi_{y\eta\eta\eta}(\tau)\eta^3. 
 \label{eq:sg-xi-perp}
  \end{gather}
 Substituting these functions to Eqs. (\ref{eq:sg-mot-x1}--\ref{eq:sg-mot-y1}) we obtain 
 ordinary differential equations:
 \begin{gather}
 \label{eq:sg-Xixyyyy1}
\frac{d}{d\tau} \left(\gamma^{(0)} \frac{d\, \Xi_{x\eta\eta\eta\eta}}{d\tau}\right) = -\frac{1}{l_y^4},\\
\label{eq:sg-Xiyyyy1}
\frac{d}{d\tau} \left(\frac{1}{\gamma^{(0)}} \frac{d\, \Xi_{y\eta\eta\eta}}{d\tau}\right) = 
-4\, \Xi_{x\eta\eta\eta\eta},\\
\label{eq:sg-Xixyy1}
\frac{d}{d\tau} \left(\gamma^{(0)} \frac{d\, \Xi_{x\eta\eta}}{d\tau}\right) = 3 \, \Xi_{y\eta\eta\eta},\\
\frac{d}{d\tau} \left(\frac{1}{\gamma^{(0)}} \frac{d\, \Xi_{y\eta}}{d\tau}\right) = 
-2 \, \Xi_{x\eta\eta},\\
\frac{d}{d\tau} \left(\gamma^{(0)} \frac{d\, \Xi_{x}}{d\tau}\right) = \Xi_{y\eta}.\label{eq:sg-Xix1}
 \end{gather}

\begin{figure}
\centering
\includegraphics[width=6 cm]{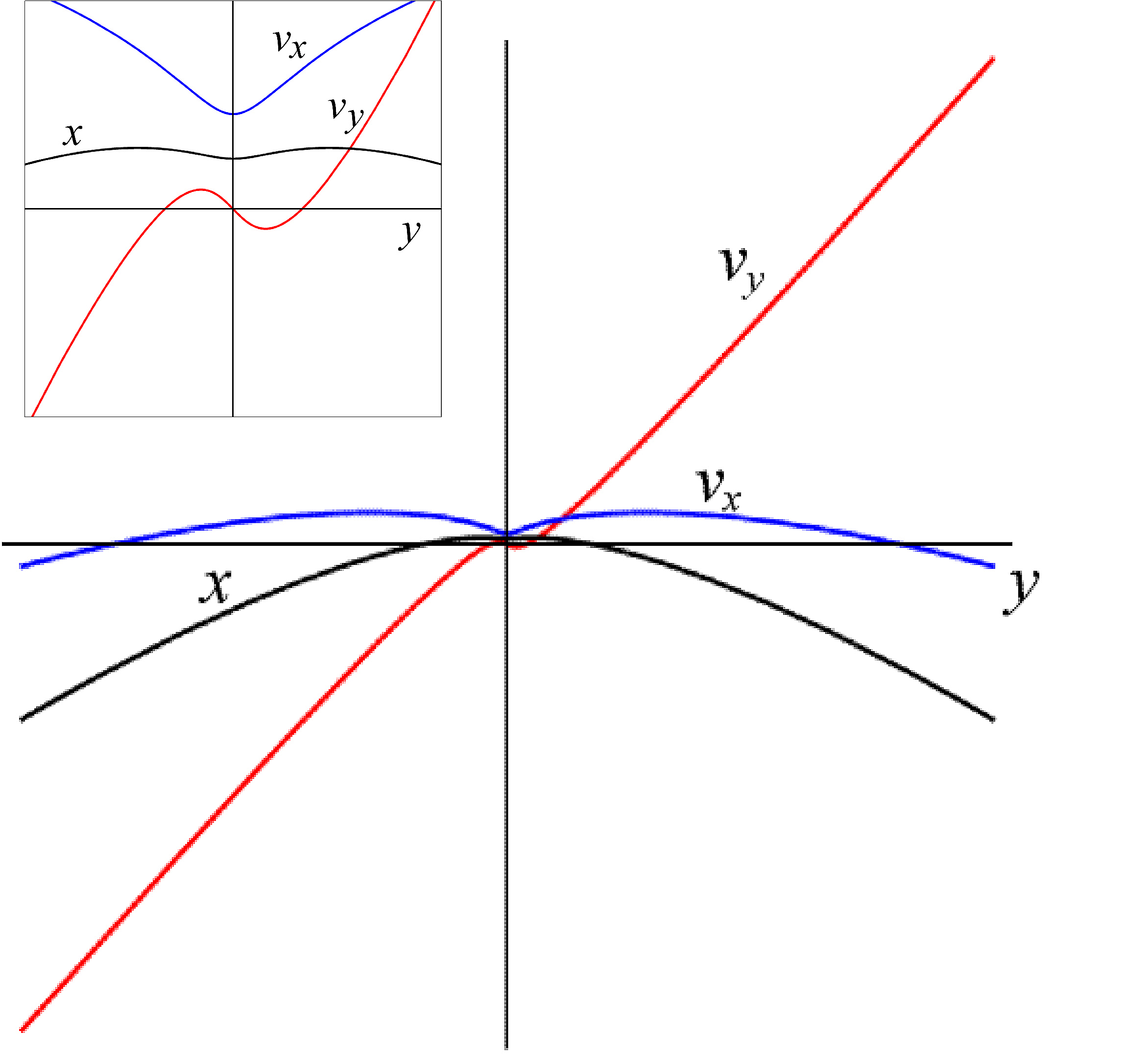}
\caption{Thin target deformation by the super-Gaussian laser pulse. The curves $x$, $v_x$ and $v_y$ show 
the target position and $x$- and $y$-components of the target element velocity in the $x,y$-plane 
for $l_y/ct_{1/3}=2$ at $t/t_{1/3}=0.5$. Inset: close-up of the near-axis region.}
\label{yVxVyX}
\end{figure}

For zero initial conditions for the displacement $\xi_i^{(1)}(\eta,0)=0$ and its time derivative 
$\dot\xi_i^{(1)}(\eta,0)=0$, 
solution to Eqs. (\ref{eq:sg-Xixyyyy1}--\ref{eq:sg-Xix1}) reads
 \begin{gather} 
 \Xi_{x\eta\eta\eta\eta}=-\frac{9 c t}{l_y^4}\left(\frac{t_{1/3}}{t}\right)^{2/3}, \,
 \Xi_{y\eta\eta\eta}=\frac{81 (c t)^2}{2 \, l_y^4}\left(\frac{t_{1/3}}{t}\right)^{2/3},\label{eq:sg-sol1}\\
 \Xi_{x\eta\eta}=\frac{4374 (c t)^3}{100 \, l_y^4}\left(\frac{t_{1/3}}{t}\right)^{4/3}, \,
 \Xi_{y\eta}=-\frac{6561 (c t)^4}{400\, l_y^4}\left(\frac{t_{1/3}}{t}\right)^{4/3}, \label{eq:sg-sol2}\\
 \Xi_{x}=-\frac{285768 (c t)^5}{156800 \, l_y^4}\left(\frac{t_{1/3}}{t}\right)^{2} \label{eq:sg-sol3}.
 \end{gather}
As it follows from expressions (\ref{eq:sg-mot-x1}--\ref{eq:sg-mot-y1}) and (\ref{eq:sg-sol1}--\ref{eq:sg-sol3}) 
the target is deformed in such the way that the periphery expands and the near-axis region contracts. This 
paradoxical behaviour can be explained by the fact that due to the density decreasing in the 
peripheral regions  the target elements there move forward faster modulating the foil curvature, which 
results in contraction of the near-axis elements, which is distinctly seen in Fig. \ref{yVxVyX}.
The longitudinal, along the $x$-axis velocity has two maxima, the transverse, $y$-component velocity gradient 
is positive at large $y$, which corresponds to the foil expansion, and it is negative near the axis 
corresponding to the 
foil compression.

\section{Results of Particle-In-Sell Simulations}

Theoretical analysis of the target off-axis displacement effects 
has been carried within the framework of the linearized model equations (\ref{eq:mot-x1}--\ref{eq:mot-z1}). 
In order to take into account the nonlinear and kinetic effects, the target deformation and instability 
we have conducted a series of 2D-PIC simulations using the two-dimensional version of relativistic 
electromagnetic code REMP \cite{REMP}. 

\begin{figure}
\centering
\includegraphics[width=9 cm]{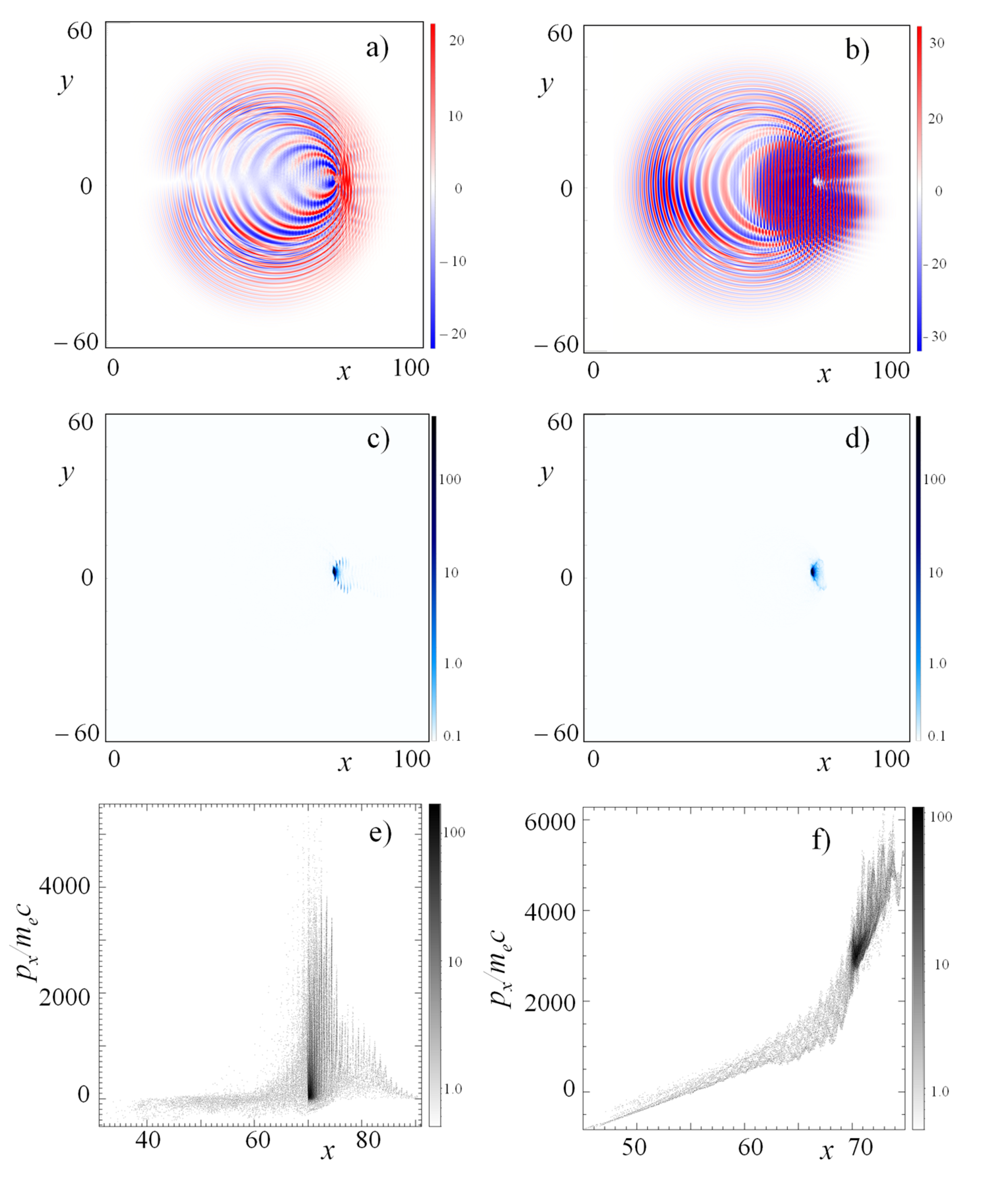}
\caption{a) and b) Distribution of the $x$- and $z$- components of the electric field; 
c) and d) of the electron and ion density in the $(x,y)$ plane; e) and f) phase planes $(x,p_x)$ of the 
electrons and ion ions, respectively, at $t=100$. The initial off-axis displacement equals $\delta y=0.25$.}
\label{fig:t100a}
\end{figure}

The simulation box is $ 300\lambda \times 100\lambda $ 
with mesh resolution of 20 cells per wavelength. 
The total number of particles is equal to $7 \times 10^4$. 
The target has the form of an ellipsoid in the $(x,y)$ plane with horizontal 
and vertical semiaxes equal to $1\lambda$ and $3.5\lambda$. 
It is initially located at $x=50\lambda$ in the near axis region with its $y$-coordinate 
varying from $0 \lambda$ to $1.8 \lambda$. 
The target comprises of hydrogen plasma with proton-to-electron mass ratio equal to 1836. 
The electron density corresponds to the ratio $\omega_{pe} /\omega = 10$. 
A circularly polarized laser pulse is excited in the vacuum region at 
the left-hand side of the computation domain. The laser pulse 
has a Gaussian shape with a length of $l_x = 20 \lambda$ and $l_y = 25 \lambda$, 
and with dimensionless amplitude $a=eE/m_e\omega c$ varies from to 250 to 325. 
Under the simulation conditions, the accelerated ion energy according to Eq. (\ref{eq:gamma(0)})
is equal to 4.5 GeV. The acceleration length $l_{acc}=ct_{acc}$ is equal to 135$\lambda$.

The aim of the PIC simulations is to investigate the dependence of the energy 
of accelerated ions on the initial displacement of the target along the $y$-axis. 

In Figs. \ref{fig:t100a} a) - d) we present electromagnetic field and electron and 
ion density distribution in the $(x,y)$ plane at $t=100$. Here and below the laser 
period $2 \pi/\omega$ and wavelength $\lambda$ are time and space units. Fig. \ref{fig:t100a} b),
with the distribution of the $z$-component of the electromagnetic field in the $(x,y)$ plane, 
shows the laser pulse reflection from the receding with relativistic velocity target. Due to 
the double Doppler effect the wavelength of the reflected light is substantially longer than 
the wavelength of the incident radiation. The laser field interaction with the plasma target is
accompanied by the high order harmonics radiation distinctly seen in the short-wavelength scattered 
radiation. The up-down asymmetry of $E_x(x,y)$ appears due asymmetry of the initial position of the target 
with respect to the laser pulse axis. There also it can be 
seen a strong longitudinal quasistatic (long wavelength) electric field formed at the rear side 
of the target. In this field the positively charged ion acceleration occurs. As it follows from 
Figs. \ref{fig:t100a} c) and d), where the electron and ion density distribution in the $(x,y)$ plane 
is shown, the electrons pushed forward by the laser radiation move almost together with the ions pulled by 
the electric charge separation electric field. In Figs. e) and f) we present the phase planes $(x,p_x)$ of the 
electrons and ion ions, respectively, which demonstrate that the highest energy electrons and ions are localised 
in the same region.

\begin{figure}
\centering
\includegraphics[width=9 cm]{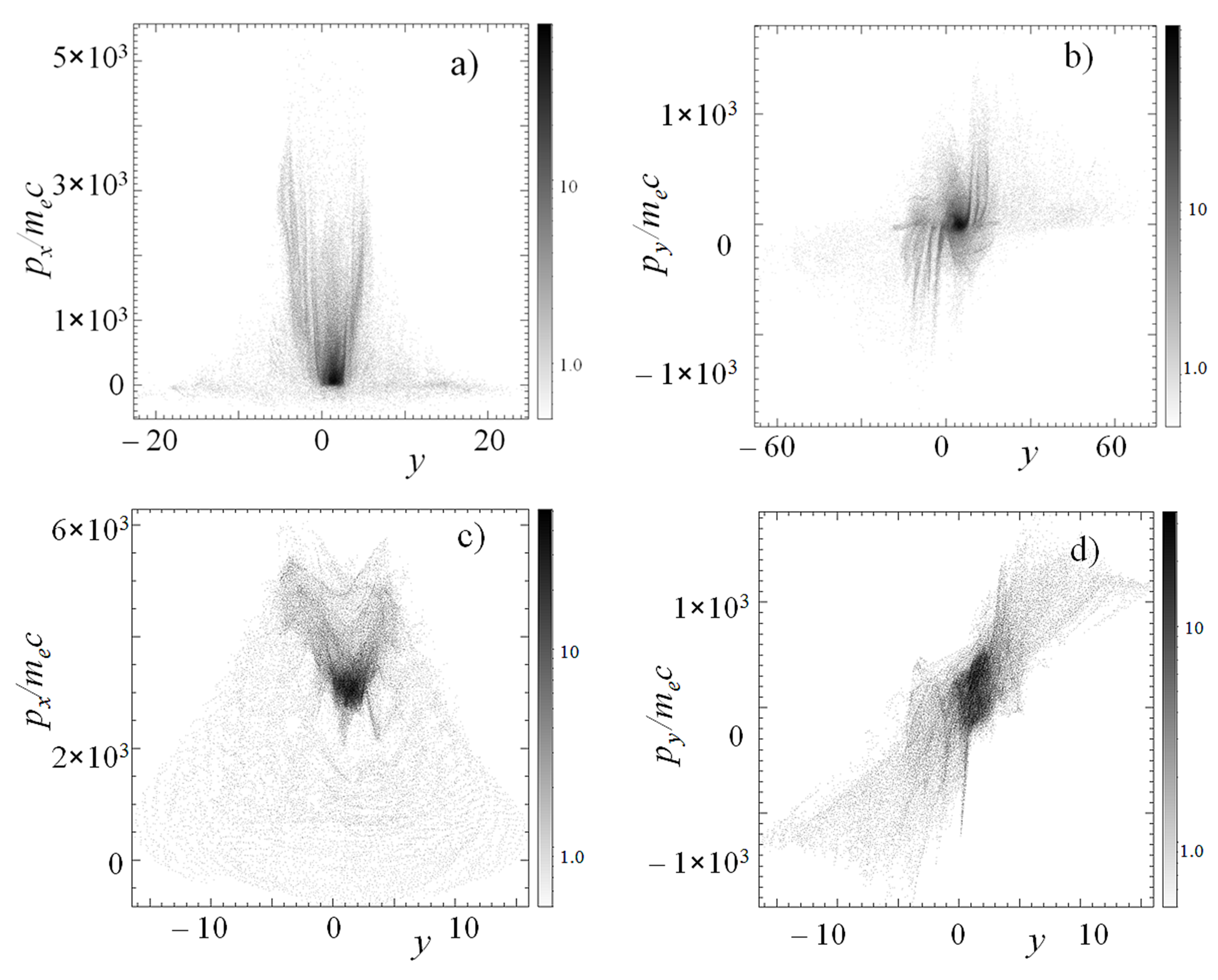}
\caption{a) and b)  Electron phase planes $(x,p_x)$ and $(y, p_y)$ and c) and d) of the 
 ions, at $t=100$ for initial $y$ coordinate 
equal to $0.25$.}
\label{fig:t100-p-1-2}
\end{figure}

In the process of nonlinear interaction with the MLT the laser pulse becomes modulated in the transverse direction 
as we can see in Fig. \ref{fig:t100a} b).
This makes the interaction with the target of initially Gaussian pulse to be similar to that of the super-Gaussian 
pulse.  As a result, the dependences of the $x$- and $y$-components of the ion and electron momentum 
on the $y$-coordinate shown in Fig. \ref{fig:t100-p-1-2} are in qualitative agreement 
with theoretical curves in Fig.\ref{yVxVyX}. Here it is possible to see a characteristic 
double maximum profile in the 
ion distribution in the $(y, p_x)$ plane. The $(y, p_y)$ distribution clearly shows the target expansion 
at its perifery and the contraction in the near-axis region.

\begin{figure}
\centering
\includegraphics[width=9 cm]{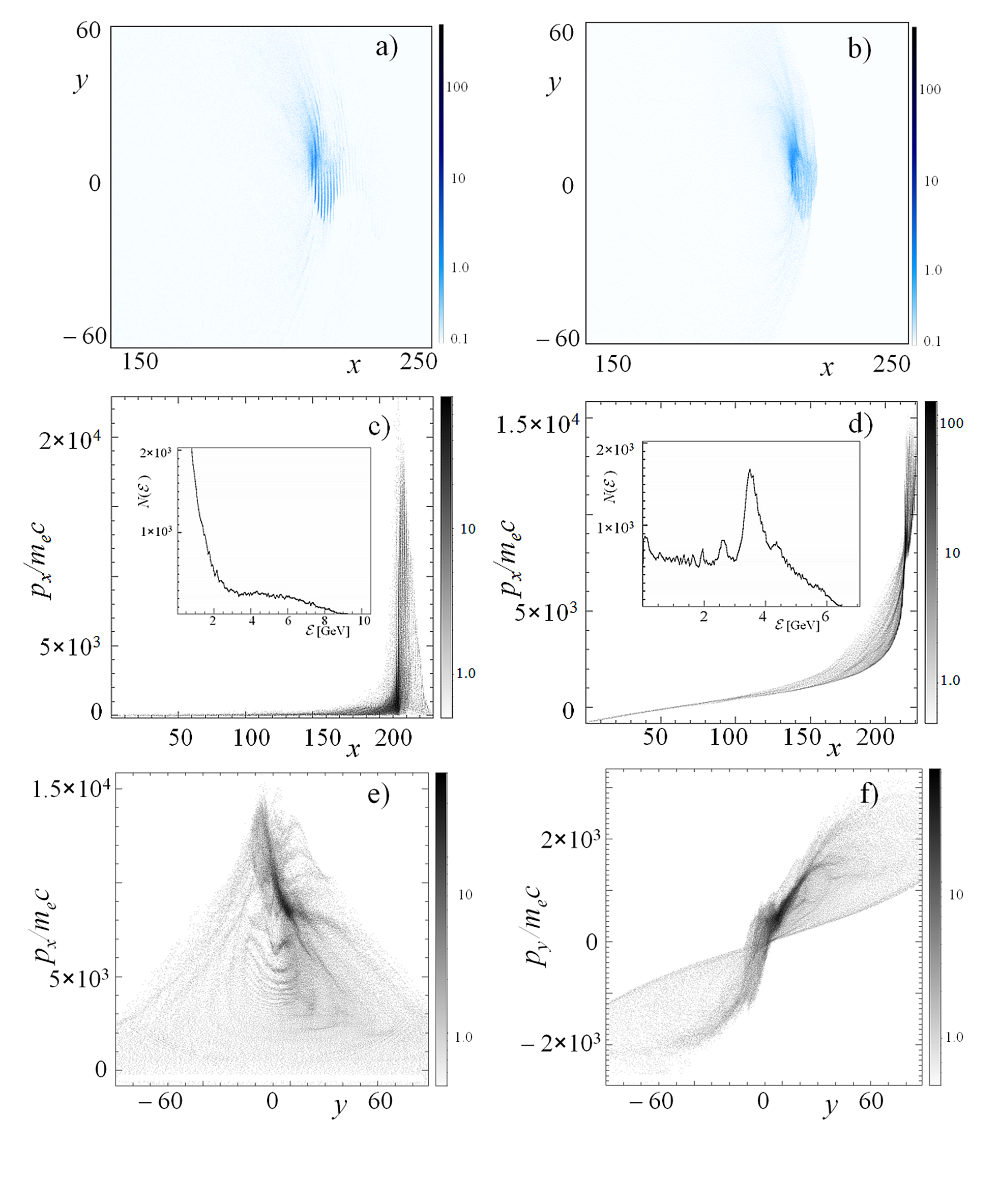}
\caption{a) and b) Distribution of the electron and ion density in the $(x,y)$ plane; 
c) and d) phase planes $(x,p_x)$ of the 
electrons and ion ions (the insets show the electron and ion energy spectra), respectively;
e) ion phase plane  $(y, p_x)$; f) ion phase plane  $(y, p_y)$, at $t=250$ for initial $y$ coordinate 
equal to $0.25$.}
\label{fig:t250}
\end{figure}

In Figs. \ref{fig:t250} a), b), where we plot distribution of the electron and ion density in the $(x,y)$ plane
 at $t=250$, 
we see, although the target is strongly deformed and displaced in the vertical direction, the ions and electrons are 
mostly localized in the same region. The c) and d) frames present the phase planes $(x,p_x)$ 
of the electrons and ion ions with the insets showing the electron and ion energy spectra, respectively. 
The electron component has a flat energy distribution with the maximum energy of the order of 8 GeV. The 
accelerated ion energy distribution shows a relatively narrow, approximately of $ 20 \%$, peak at the energy of the 
order of 4 GeV. The ion phase plane  $(y, p_x)$ and  ion phase plane  $(y, p_y)$ in 
Figs. \ref{fig:t250} e) and f) demonstrate that the high energy ions remain localized in the near-axis region.

\begin{figure}
\centering
\includegraphics[width=8 cm]{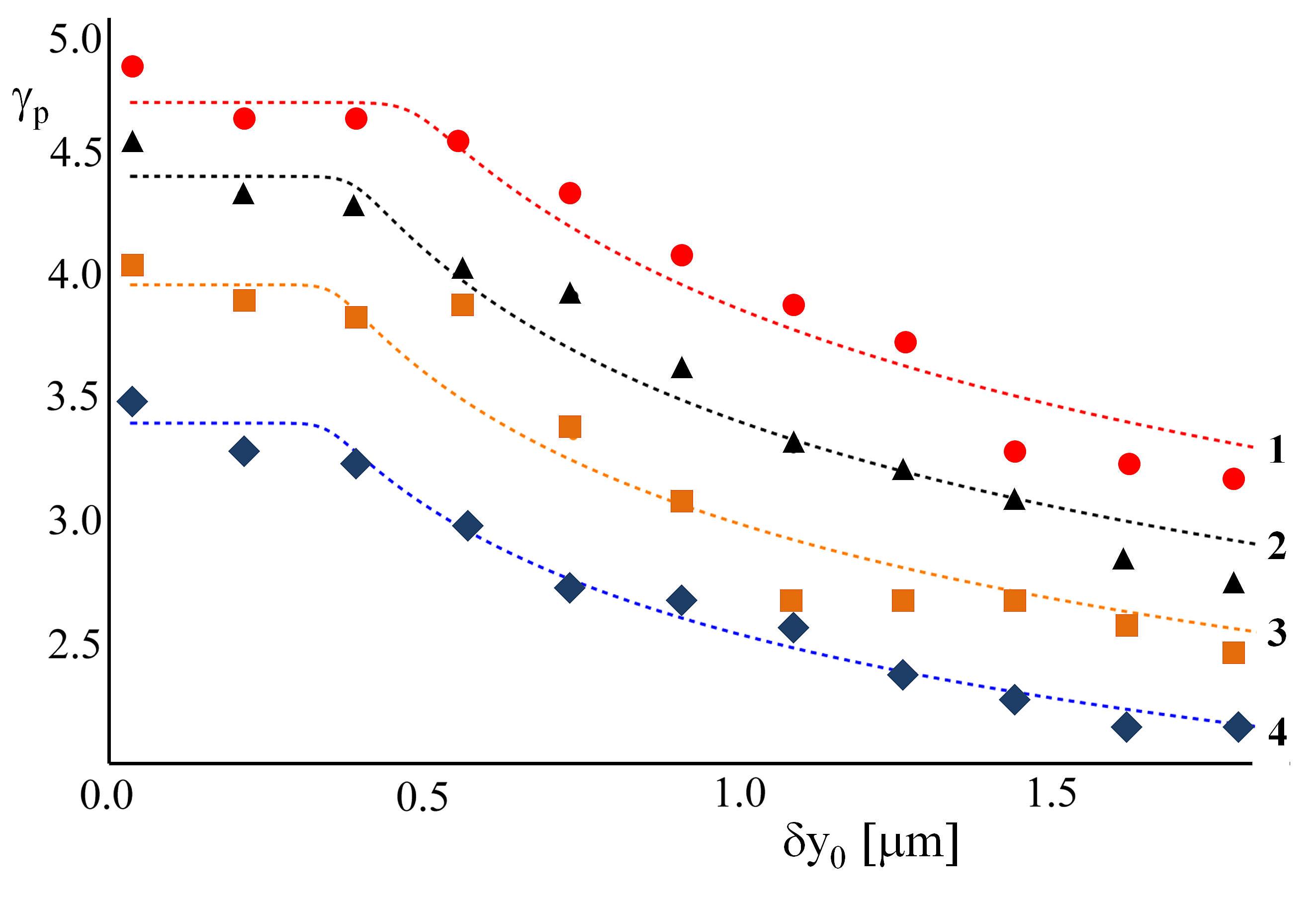}
\caption{Normalized proton energy gained with the MLT initially shifted in the vertical direction 
versus the initial target coordinate for different amplitudes of the Gaussian laser pulse 
(1. $a=325$; 2. $a=300$; 3. $a=275$; 4. $a=250$). 
The plot markers are the simulation results and the curves correspond to theoretical dependences.}
\label{fig:MaxEn}
\end{figure}

Dependence of the accelerated ion energy, ${\cal E}_{\alpha}$, on the target initial position, $\delta r_0$, is presented in 
Fig. \ref{fig:MaxEn} for different laser pulse amplitude. 
At the simulation conditions the accelerated ions reach their maximum energy at the 
time approximately equal to 250 fs, so all of the graphs are presented at that moment of time.
Here we plot the theoretical curves (dashed lines) calculated by using Eqs. (\ref{eq:tau-perp}--\ref{eq:t-1/3}) 
and the energy value obtained in simulations (dots in color). 
The theoretical dependence of the ion energy on the inititial target position follows 
from Eqs. (\ref{eq:tau-perp}) and (\ref{eq:gamma-perp}). It reads
 \begin{equation}  
\label{eq:E-dr0}
{\cal E}_{\alpha,\delta r_0}
=m_{\alpha}c^2\frac{2^{1/2}}{3}\frac{l_{\perp}^{3/4}}{c^{1/2}t_{1/3}^{1/2} \delta r_0^{1/4}}.
\end{equation}
This expression is valid in the limit of substantially large $\delta r_0$. 
When $\delta r_0 \to 0$ it formally tends to infinity. Apparently, the ion energy from the off-axis localized 
target cannot be larger that the ion energy in the case of the target positioned exactly on the axis, 
${\cal E}_{\alpha, max}$. In order to take this into account we shall use the  interpolation formula
 \begin{equation}  
\label{eq:E-interpol}
\frac{1}{{\cal E}_{\alpha}^{s}}=\frac{1}{{\cal E}_{\alpha, max}^{s}}+\frac{1}{{\cal E}_{\alpha, \delta r_0}^{s}}
\end{equation}
with the fitting parameter $s >>1$. In the limit of small $\delta r_0$ the ion energy is equal 
to ${\cal E}_{\alpha, max}$. For large initial vertical coordinate it is proportional to $\delta r_0^{-1/4}$
 accortding to Eq. (\ref{eq:E-dr0}).
In Fig. \ref{fig:MaxEn} we plot the normalized ion energy $\gamma_p$ achieved 
with the MLT initially shifted in the vertical direction 
versus the initial target coordinate $\delta y_0$ for different amplitudes of the Gaussian laser pulse 
The plot markers are the 2D PIC simulation results and the curves correspond to theoretical dependences given 
by Eq. (\ref{eq:E-interpol}) for 1. $a=325$; 2. $a=300$; 3. $a=275$; 4. $a=250$).  

For small the initial target coordinate the ion energy decreases with $\delta y_0$ 
more slowly than it is predicted by the theory due to self-modulation of the 
laser pulse in transverse direction, which is distinctly seen in the electromagnetic field didstribution 
in Fig. \ref{fig:t100a} b). The laser pulse self-modulation prevents the target from sleapage out off the acceleration
phase providing the fast ion collimation seen in Fig. \ref{fig:t250} e). As it follows from dependences presented 
in Fig. \ref{fig:MaxEn}, the laser pulse modulation effects are significant for $\delta y_0 <0.5 \mu$m.

\section{Conclusions and discussions}

We have studied the effects of the laser pulse transverse inhomogeneity on the ion acceleration 
in the RPDA regime. Within the framework of a thin foil approximation we found the dependence 
of the accelerated ion maximum energy on the off-axis displacement of the mass limited target 
for Gaussian and super-Gaussian laser pulse profiles. 
When the target is irradiated by the Gaussian laser pulse it is pushed away from the pulse 
by the ponderomotive pressure of electromagnetic radiation, 
while in the case super-Gaussian the central part of the target may undergo self-contraction 
provided its initial of-axis displacement 
is small enough. The 2D particle in cell simulations affirm the theoretical calculations at large 
 initial coordinate of the target in the vertical direction,  $\delta y_0$. 
If the target is positioned in the vicinity of the axis, the self-modulation of the 
laser pulse in transverse direction prevents the target from sleapage out off the acceleration
phase providing the fast ion collimation.

The results obtained can be used for determining 
the required laser-target alignment parameters  
and/or diagnostics of the ion acceleration 
by the laser radiation pressure with mass limited targets, 
widely used in the experiments.



\bibliographystyle{unsrt} 

\begin{thebibliography}{99}

\bibitem{IonsB} M. Borghesi, J. Fuchs, S. V. Bulanov, A. J. Mackinnon, P. Patel,  and M. Roth, 
{\it Fus. Sci. and Technology} {\bf 49}, 412 (2006)

\bibitem{IonsAVK} A. V. Korzhimanov, A. A. Gonoskov, E. A. Khazanov, and A. M. Sergeev, {\it Phys. Usp.} {\bf 54}, 9 (2011)

\bibitem{IonsD} H. Daido, M. Nishiuchi,  and A. S. Pirozhkov, 
{\it Rep. Prog. Phys.} {\bf 75}, 056401 (2012)

\bibitem{IonsM} A. Macchi, M. Passoni,  and M. Borghesi, 
{\it Rev. Mod. Phys.} {\bf 85}, 751 (2013)

\bibitem{IonsGUS}  S. Yu. Gus'kov, {\it Plasma Phys. Rep.} {\bf 39}, 1 (2013)

\bibitem{HTUFN} S. V. Bulanov, J. J. Wilkens, M. Molls, T. Zh. Esirkepov, G. Korn, G. Kraft, S. D. Kraft,  and 
V. S. Khoroshkov, {\it Phys. Usp.} {\bf 57}, 1265 (2014)

\bibitem{RelAstro} S. V. Bulanov, T. Zh. Esirkepov, M. Kando, J. Koga, K. Kondo,  and G. Korn, {\it Plasma Phys. Rep.} {\bf 41}, 1 (2015)

\bibitem{BesBook} V. S. Beskin, {\it MHD Flows in Compact Astrophysical Objects}. (Springer, Berlin, 2010)

\bibitem{Load1} E. V. Derishev, V. V. Kocharovsky, and Vl. V. Kocharovsky {\it ApJ}, {\bf 521}, 640 (1999)
 
\bibitem{Load2} B. E. Stern  and J. Poutanen, {\it MNRAS} {\bf 383}, 1695 (2008) 

\bibitem{Prince} D. Khangulyan, F. Aharonian,  and V. Bosch-Ramon {\it MNRAS}, {\bf 383}, 467 (2008)

\bibitem{Prince} B. Cerutti, A. Philippov, K. Parfrey,  and A. Spitkovsky {\it MNRAS} (in press) http://arxiv.org/abs/1410.3757

\bibitem{TNSAG} A. V. Gurevich, L. V. Pariskaya, and L. P. Pitaevskii, 
{\it Sov. Phys. JETP} {\bf 22}, 449 (1966)

\bibitem{TNSAM} P. Mora, {\it Phys. Rev. Lett.} {\bf 90}, 185002 (2003)

\bibitem{TNSAW} S. Wilks, W. L. Kruer, M. Tabak, and A. B. Langdon, {\it Phys. Rev. Lett.} {\bf 69}, 1383 (1992)

\bibitem{TNSAH} S. P. Hatchett, C. G. Brown, T. E. Cowan, E. A. Henry, J. S. Johnson, 
M. H. Key, J. A. Koch, A. B. Langdon, B. F. Lasinski, R. W. Lee, A. J. Mackinnon, 
D. M. Pennington, M. D. Perry, T. W. Phillips, M. Roth, T. C. Sangster, M. S. Singh, 
R. A. Snavely, M. A. Stoyer, S. C. Wilks, and K. Yasuike, 
{\it Phys. Plasmas} {\bf 7}, 2076 (2000)

\bibitem{CEL} I. Last, I. Schek, and J. Jortner, {\it J. Chem. Phys.} {\bf 107}, 6685 (1997)

\bibitem{CEN} K. Nishihara, H. Amitani, M. Murakami, S. V. Bulanov,  and T. Zh. Esirkepov, {\it Nucl. Instrum. Meth. Phys. Res. A} {\bf  464}, 98 (2001)

\bibitem{CEB} V. F. Kovalev and V. Yu. Bychenkov, {\it Phys. Rev. Lett.} {\bf 90}, 185004 (2003)

\bibitem{MM} M. Murakami and M. M. Basko, {\it Phys. Plasmas} {\bf 13}, 012105 (2006)

\bibitem{RPDAV} V. I. Veksler, {\it At. Energ.} {\bf 2}, 427 (1957)

\bibitem{RPDAE} T. Zh. Esirkepov, M. Borghesi, S. V. Bulanov, G. Mourou, and T. Tajima, {\it Phys. Rev. Lett.} {\bf 92}, 175003 (2004)

\bibitem{RPDAK} O. Klimo, J. Psikal, J. Limpouch, and V. T. Tikhonchuk, 
{\it Phys. Rev. ST Accel. Beams} {\bf 11}, 031301 (2008)

\bibitem{RPDAR} A. P. L. Robinson,  M. Zepf, S. Kar, R. G. Evans, and C. Bellei, {\it New J. Phys.} {\bf 10}, 013021 (2008)

\bibitem{FOILV} A. V. Vshivkov, N. M. Naumova, F. Pegoraro, and S. V. Bulanov, {\it Phys. Plasmas} {\bf 5}, 2752 (1998)

\bibitem{FOILSVB} S. V. Bulanov, T. Zh. Esirkepov, M. Kando, S. S. Bulanov,
S. G. Rykovanov, and F. Pegoraro, {\it Phys. Plasmas} {\bf 20}, 123114 (2013)

\bibitem{RPDASS} S. S. Bulanov, C. B. Schroeder, E. Esarey, and W. P. Leemans, 
{\it Phys. Plasmas} {\bf 19}, 093112 (2012)

\bibitem{UFNMIRR} S. V. Bulanov, T. Zh. Esirkepov, M. Kando, A. S. Pirozhkov, N. N. Rosanov, 
{\it Phys. Usp.} {\bf 56}, 429 (2013)


\bibitem{SKar1} S. Kar, M. Borghesi, S. V. Bulanov, A. Macchi, M. H. Key, T. V.
 Liseykina, A. J. Mackinnon, P. K. Patel, L. Romagnani, A. Schiavi, and O. Willi, 
{\it Phys. Rev. Lett.} {\bf 100}, 225004 (2008)

\bibitem{Henig} A. Henig, S. Steinke, M. Schnuerer, T. Sokollik, R. Hoerlein, D. Kiefer, D. Jung, J. Schreiber, 
B. M. Hegelich, X. Q. Yan, J. Meyer-ter-Vehn, T. Tajima, P. V. Nickles, W. Sandner, and D. Habs,
{\it Phys. Rev. Lett.} {\bf 103}, 245003 (2009)

\bibitem{SKar2} S. Kar, K. F. Kakolee, B. Qiao, A. Macchi, M. Cerchez, D. Doria, M. Geissler, P. McKenna, D.
 Neely, J. Osterholz, R. Prasad, K. Quinn, B. Ramakrishna, G. Sarri, O. Willi, X. Y. Yuan, M. Zepf, and M. Borghesi, 
{\it Phys. Rev. Lett.} {\bf 109}, 185006 (2012)

\bibitem{JLIM} J. Limpouch, J. Psikal, A.A. Andreev, K. Yu. Platonov, and S. Kawata, {\it Laser and Particle Beams}
 {\bf 26}, 225 (2008)

\bibitem{AAA} A. A. Andreev, J. Limpouch, J. Psikal, K. Yu. Platonov, and V. T. Tikhonchuk, 
{\it Eur. Phys. J. ST} {\bf 175}, 123 (2009) 

\bibitem{Kluge} T. Kluge, W. Enghardt, S. D. Kraft, U. Schramm, K. Zeil, T. E. Cowan and M. Bussmann,
 {\it Phys. Plasmas} {\bf 17}, 123103 (2010)

\bibitem{Zeil} K. Zeil, J. Metzkes, T. Kluge, M. Bussmann, T. E Cowan, 
S. D. Kraft, R. Sauerbrey, B. Schmidt, M. Zier, and U. Schramm,
{\it Plasma Phys. Control. Fusion} {\bf 56}, 084004 (2014) 

\bibitem{AZ} A. Zigler, S. Eisenman, M. Botton, E. Nahum, E. Schleifer, A. Baspaly, I. Pomerantz, F. Abicht, J. Branzel, G. Priebe, S. Steinke, 
A. Andreev, M. Schnuerer, W. Sandner, D. Gordon, P. Sprangle, and K. W. D. Ledingham, {\it Phys. Rev. Lett.} {\bf 110}, 215004 (2013)

\bibitem{Wang} J. W. Wang, M. Murakami, S. M. Weng, H. Xu, J. J. Ju, S. X. Luan, and W. Yu, {\it Phys. Plasmas} {\bf 21}, 123103 (2014). 

\bibitem{Yu} T. P. Yu, Z. M. Sheng, Y. Yin, H. B. Zhuo, Y. Y. Ma, F. Q. Shao and A. Pukhov
{\it Phys. Plasmas} {\bf 21}, 053105 (2014)

\bibitem{Fukuda2009} Y. Fukuda, A. Ya. Faenov, M. Tampo, T. A. Pikuz, T. Nakamura, M. Kando, Y. Hayashi, A. Yogo, H. Sakaki, 
 T. Kameshima, A. S. Pirozhkov, K. Ogura, M. Mori, T. Zh. Esirkepov, J. Koga, A. S. Boldarev, V. A. Gasilov, 
 A. I. Magunov, T. Yamauchi, R. Kodama, P. R. Bolton, Y. Kato, T. Tajima, H. Daido, and S. V. Bulanov, 
{\it Phys. Rev. Lett.} {\bf 103}, 165002 (2009)

\bibitem{YuPRL} T.-P. Yu, A. M. Pukhov, Z.-M. Sheng, F. Liu, and G. Shvets, {\it Phys. Rev. Lett.} {\bf 110}, 045001 (2013)

\bibitem{UNLIM} S. V. Bulanov,   E. Yu. Echkina,  T. Zh. Esirkepov, I. N. Inovenkov, M. Kando, F. Pegoraro, 
and G. Korn, 
{\it Phys. Rev. Lett.} {\bf 104}, 135003 (2010)

\bibitem{UNLIMPoP} S. V. Bulanov,   E. Yu. Echkina,  T. Zh. Esirkepov, I. N. Inovenkov, M. Kando, F. Pegoraro, 
and G. Korn, 
{\it Phys. Plasmas} {\bf 17}, 063102 (2010)


\bibitem{PhotonBubbles} F. Pegoraro and S. V. Bulanov, {\it Phys. Rev. Lett.} {\bf 99}, 065002 (2007)

\bibitem{REMP}  T. Zh. Esirkepov, {\it Comput. Phys. Commun.} {\bf 135}, 144 (2001)

\bibitem{Ott} E. Ott, {\it Phys. Rev. Lett.} {\bf 29}, 142 (1972)

\bibitem{gen1}  W. Manheimer, D. Colombait, and E. Ott, 
{\it Phys. Fluids} {\bf 27}, 2164 (1984)

\bibitem{gen2} T. Taguchi and K. Mima, {\it Phys. Plasmas} {\bf 2}, 2790 (1995) 


\bibitem{KndK} G. A. Korn and T. M. Korn, {\it Mathematical Handbook for
Scientists and Engineers.} (Dover Publ., New York, 2000)

\bibitem{Ech10} E. Yu. Echkina, 
 I. N. Inovenkov, T. Zh. Esirkepov, F. Pegoraro, M. Borghesi, and S. V. Bulanov,
{\it Plasma Phys. Rep.} {\bf 36}, 15 (2010)

\bibitem{Slow} S. V. Bulanov, T. Zh. Esirkepov, M. Kando, F. Pegoraro, S. S. Bulanov, 
C. G. R. Geddes, C. B. Schroeder, E. Esarey, and W. P. Leemans,
{\it Phys. Plasmas} {\bf 19}, 103105 (2012)

\bibitem{OPTAM} A. Macchi, S. Veghini, and F. Pegoraro, 
{\it Phys. Rev. Lett.} {\bf 103}, 085003 (2009)

\bibitem{Humphries} S. Humphries, Jr., {\it Principles of Charged Particle Acceleration} (Wiley, New York, 1999) 

\end{thebibliography}

\end{document}